\documentclass[aip,rsi,reprint]{revtex4-1}
\usepackage{indentfirst,amsmath,amssymb}
\usepackage{graphicx}
\usepackage[utf8]{inputenc}

\begin{document}
 \title{Enhanced interaction between a mechanical oscillator and two coupled resonant electrical circuits}
 \author{A. V. Dmitriev}\email{dmitriev@hbar.phys.msu.ru}
 \author{V. P. Mitrofanov}\email{mitr@hbar.phys.msu.ru}
 \affiliation{Faculty of Physics, Lomonosov Moscow State University, Moscow 119991, Russia}
 \begin{abstract}
This paper reports result of calculation and experimental realization of an electromechanical system that consists of a high-Q mechanical oscillator parametrically coupled in the manner of a capacitive transducer with a RF circuit, which is in turn inductively coupled with another RF circuit. The system operates in the resolved sideband regime when the mechanical oscillator's frequency is larger than the electrical circuits' bandwidths. Using two coupled RF circuits allowed one to enhance the interaction between them and the mechanical oscillator which is one of flexural vibrational modes of a free-edge circular silicon wafer. Such a coupled electromechanical system can be used as a high-sensitive capacitive vibration sensor.
 \end{abstract}
 \maketitle
 \section{Introduction}
  Cold damping, i.e. cooling of mechanical oscillators by pumping out their energy into optical or microwave cavity, is widely used to prepare nanomechanical oscillators near the ground quantum state \cite{Rocheleau2010,Teufel2011,Chan2011}. The insertion of additional damping into a mechanical oscillator using capacitive coupling between the oscillator and a resonant radio frequency (RF) circuit was firstly demonstrated in Ref.~\onlinecite{Braginsky1977}. Dynamic back action of the parametric capacitive transducer resulted in increasing or decreasing $Q$-factor of the mechanical oscillator. In such systems the circuit bandwidth was much larger than the mechanical resonant frequency. In modern optomechanical systems additional damping is inserted into a mechanical oscillator by coupling it parametrically with a pumped optical cavity whose resonant frequency is detuned from the pumping frequency by the value of mechanical resonant frequency. If the cavity bandwidth is much smaller than the mechanical resonant 
frequency the so-called resolved sideband damping regime occurs \cite{Kippenberg2007,Schliesser2008}. This way of inserting additional damping exploits the resonance amplification of the anti-Stokes sideband while the Stokes sideband is small because it lies outside the bandwidth of the cavity.
 
  This effect can be enhanced by coupling the mechanical oscillator to a system that consists of two coupled optical cavities or resonant RF circuits with frequencies which are split exactly by the value of the mechanical oscillator frequency. In this case the signal at the driving frequency and its anti-Stokes sideband can be resonantly amplified simultaneously. At present the three-mode systems formed by two optical modes and a mechanical oscillator are of major interest in optomechanics \cite{Kippenberg2010,Fan2012,Liu2014}. In particular, they are studied under the perspective of their potential applications as macroscopic quantum systems \cite{Ludwig2012,Aggarwal2014}. 

 In this paper, we present results of calculation and experimental realization of an electromechanical system in which the mechanical oscillator interacts with two coupled resonant RF circuits in the resolved sideband regime. This allows us to enhance the interaction between the mechanical oscillator and resonant RF circuits. The mechanical oscillator is one of flexural vibration modes of a free-edge circular silicon wafer \cite{Dmitriev2014}.
 
 \section{Theoretical analysis}
\begin{figure}
 \includegraphics[width=\columnwidth]{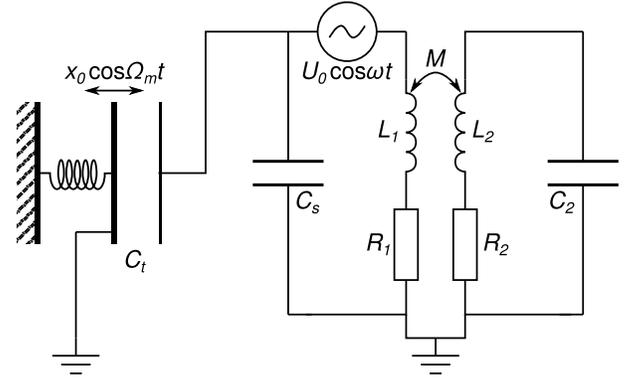}
 \caption{Schematic of the three-mode system formed by mechanical oscillator and two coupled resonant electrical circuits.\label{theor-scheme}}
\end{figure}

 Schematic of the considered system is shown in Fig.~\ref{theor-scheme}. A mechanical oscillator with lumped parameters: mass $m$, angular resonant frequency $\Omega_m$ and damping rate $\Gamma_m=\Omega_m/Q_m$ is parametrically coupled with a RF circuit that is formed by capacitance $C_{1T}$, inductance $L_1$ and resistance $R_1$ and has angular resonant frequency $\nu_1=1/\sqrt{C_{1T}L_1}$. This circuit is pumped with RF voltage source $U = 1/2(U_0 e^{i\omega t}+c.c.)$
 
 The total capacitance of this RF circuit $C_{1T}$ may be represented as a sum of the capacitance of the parallel-plate capacitive transducer $C_t$ and stray capacitance $C_s$. The capacitance $C_t$ between the rigidly mounted transducer plate and the moving plate mounted on the oscillating mass is modulated by the motion of the mechanical oscillator $x(t)$ and can be written down in the form $C_t=C_t^{(0)}/(1-x(t)/d)$, where $C_t^{(0)}=\varepsilon_0 S/d$ and $d$ are respectively the capacitance of the transducer and the gap between the plates in the absence of mechanical motion, $\varepsilon_0$ is the vacuum dielectric constant and $S$ is the plate area.
 
 Since the mechanical displacement is small as compared with the capacitor gap $d$ one can introduce a small parameter $\alpha=x_0/d << 1$ so that $x=x_0\xi(t)=\alpha d \xi(t)$, where $|\xi(t)|\le 1$. In this case the capacitance of the primary circuit can be written down with the accuracy of $\alpha^2$ as
 $$C_{1T}\simeq\frac{C_1}{1-\beta\alpha\xi(t)},$$ where $C_1 = C_t^{(0)}+C_s$ and $\beta = C_t^{(0)}/(C_t^{(0)}+C_s)$.
 
 The electric charge $q_t$ on the plates of the transducer capacitor $C_t$ produces the attractive force between the plates $F=q_t^2/(2\varepsilon_0 S)$.
 
 This (primary) RF circuit is inductively coupled with another (secondary) RF circuit that has angular resonant frequency $\nu_2$, capacitance $C_2$, inductance $L_2$, resistance $R_2$. The mutual inductance between the circuits is $M$.

 The governing equations for the considered system can be written in the following form:
\begin{alignat}{1}
 &\ddot q_1 + \Gamma_1 \dot q_1+\nu_1^2 q_1+\kappa_1\ddot q_2 - \frac{\nu_1^2 \beta}{d} q_1 x = \frac 12 u_0 e^{i \omega t}+c.c.\label{gov-q1}\\
 &\ddot q_2 + \Gamma_2 \dot q_2+\nu_2^2 q_2+\kappa_2\ddot q_1 = 0,\label{gov-q2}\\
 &\ddot x+\Gamma_m \dot x + \Omega_m^2 x = \frac{\beta^2 q_1^2}{2\varepsilon_0 S m},\label{gov-x}
\end{alignat}
 where $q_1$ and $q_2$ are the values of the electric charge on the capacitors $C_{1T}$ and $C_2$, respectively, $\kappa_{1,2}=M/L_{1,2}$ are the coupling factors, $\Gamma_{1,2}=R_{1,2}/L_{1,2}$ and $u_0=U_0/L_1$. 
 
 Since the parameter $\alpha$ is small ($x$ is proportional to $\alpha$), one can take the solution in the form of series expansion by $\alpha$, neglecting all terms proportional to $\alpha^n$, where $n\ge 2$:
 \begin{align}
  q_1=\frac{1}{2} \left[q_{10}+\alpha q_{11}(t)\right]e^{i\omega t}+c.c.,\label{solform-q1}\\
  q_2=\frac{1}{2} \left[q_{20}+\alpha q_{21}(t)\right]e^{i\omega t}+c.c.,\label{solform-q2}\\
  x=\frac{1}{2} \alpha d e^{i\Omega_m t}+c.c.\label{solform-x}
 \end{align}
 
 Substituting Eqs. (\ref{solform-q1}-\ref{solform-x}) to (\ref{gov-q1}-\ref{gov-x}), equaling terms with the same powers of $\alpha$ and taking into account that $q_{11}$ and $q_{21}$ can be treated as slowly varying amplitudes (i.e. neglecting the terms proportional to $\ddot q_{i1}$ and $\Gamma_i \dot q_{i1}$, $i=1,2$) one can obtain the following system of equations:
 \begin{alignat}{2}
  \left[-\omega^2+i\omega\Gamma_1 +\nu_1^2\right]q_{10}-\kappa_1 \omega^2 q_{20} &= u_0; \label{stationary-q1} \\
  \left[-\omega^2+i\omega\Gamma_2 +\nu_2^2\right]q_{20}-\kappa_2 \omega^2 q_{10} &= 0; \label{stationary-q2}
 \end{alignat}
 \begin{multline}
  2 i\omega \dot q_{11} + \left[ -\omega^2 + i\omega\Gamma_1 +\nu_1^2\right]q_{11}-\kappa_1\omega^2 q_{21} \\
  = \frac{1}{2} \beta\nu_1^2 q_{10}\left[ e^{i\Omega_m t}+e^{-i\Omega_m t}\right];\label{perturb-q1}
 \end{multline}
 \begin{equation}
  2 i\omega \dot q_{21} + \left[ -\omega^2 + i\omega\Gamma_2 +\nu_2^2\right]q_{21}-\kappa_2\omega^2 q_{11} =0.\label{perturb-q2}
 \end{equation}
 The first pair of equations describes the stationary solution for the electrical subsystem in the absence of any mechanical displacement. It gives the following expression for $q_{10}$:
 \begin{equation}
  q_{10}=u_0\omega^{-2}\chi(\omega)e^{i\Psi(\omega)}\label{sol-q10},
 \end{equation}
 where 
 \begin{alignat*}{2}
  \chi(\omega)=\frac{1}{\sqrt{\Delta_{eff}^2(\omega)+\theta_{eff}^2(\omega)}},\\
  \tan\Psi(\omega)= -\theta_{eff}(\omega)/\Delta_{eff}(\omega),
 \end{alignat*}
 $\Delta_{eff}$ and $\theta_{eff}$ are expressed as following:
 \begin{alignat}{2}
  \Delta_{eff}(\omega) =& - &\Delta_1(\omega) +& \frac{\kappa_1\kappa_2 \Delta_2(\omega)}{\Delta_2^2(\omega)+\theta_2^2(\omega)},\label{deltaeff}\\
  \theta_{eff}(\omega) =& &\theta_1(\omega) +& \frac{\kappa_1\kappa_2 \theta_2(\omega)}{\Delta_2^2(\omega)+\theta_2^2(\omega)},\label{thetaeff}
 \end{alignat}
 where $\Delta_{1,2}(\omega)=1-\nu_{1,2}^{2}/\omega^2$ are the dimensionless detuning functions, $\theta_{1,2}(\omega)=\Gamma_{1,2}/\omega$ are the dimensionless damping rates.
 
 Substituting (\ref{sol-q10}) into (\ref{perturb-q1}) one can solve the second pair of equations and get the following results:
 \begin{equation}
  q_{11}=q_{11+}e^{i\Omega_m t} + q_{11-}e^{-i\Omega_m t},\label{sol-q11}
 \end{equation}
 \begin{equation}
  q_{11\pm}=\frac{1}{2}\frac{\beta\nu_1^2 u_0}{\omega^2 (\omega\pm\Omega_m)^2}\chi(\omega)\chi(\omega\pm\Omega_m) e^{i[\Psi(\omega)+\Psi(\omega\pm\Omega_m)]}.\label{sol-q11pm}
 \end{equation}
 
 Now one can calculate the expression for the square of $q_1(t)$, which the force acting on the mechanical subsystem is proportional to:
 \begin{widetext}
 \begin{multline}
  q_1^2(t)=\frac{\alpha}{4}\beta\nu_1^2 u_0^2\omega^{-4}\chi^2(\omega)\biggl\{\left[ \frac{\chi(\omega+\Omega_m)}{(\omega+\Omega_m)^2}\cos\Psi(\omega+\Omega_m) + \frac{\chi(\omega-\Omega_m)}{(\omega-\Omega_m)^2}\cos\Psi(\omega-\Omega_m)\right] e^{i\Omega_m t} \\ +i\left[ \frac{\chi(\omega+\Omega_m)}{(\omega+\Omega_m)^2}\sin\Psi(\omega+\Omega_m) - \frac{\chi(\omega-\Omega_m)}{(\omega-\Omega_m)^2}\sin\Psi(\omega-\Omega_m)\right]e^{i\Omega_m t} \biggr\}+c.c.,
  \label{sol-q1sq}
 \end{multline}
 where the terms oscillating at frequencies of about $\omega$ as well as the terms proportional to $\alpha^2$ and the constant term are neglected because they do not make any significant impact on the oscillator's motion.
 
 The first term in the expression (\ref{sol-q1sq}) represents the in-phase component of the force acting on the oscillator; one can treat this component as one inserting additional stiffness and producing an effective shift $\Delta\Omega_m$ of the oscillator resonant frequency $\Omega_m$:
 \begin{equation}
  \Delta\Omega_m = \frac{S U_0^2 \beta^3 \varepsilon_0}{8 d^3 m} \frac{\nu_1^6}{\Omega_m\omega^4} \chi^2(\omega)\\ 
  \left[ \frac{\chi(\omega+\Omega_m)}{(\omega+\Omega_m)^2}\cos\Psi(\omega+\Omega_m) + \frac{\chi(\omega-\Omega_m)}{(\omega-\Omega_m)^2}\cos\Psi(\omega-\Omega_m) \right]\label{domega}
 \end{equation}
 
 The second term represents the quadrature force component that inserts additional damping in the oscillator generated by the transducer backaction:
 \begin{equation}
  \Delta Q_m^{-1} = \frac{S U_0^2 \beta^3 \varepsilon_0}{4 d^3 m} \frac{\nu_1^6}{\Omega_m^2\omega^4} \chi^2(\omega)\left[ \frac{\chi(\omega-\Omega_m)}{(\omega-\Omega_m)^2}\sin\Psi(\omega-\Omega_m) - \frac{\chi(\omega+\Omega_m)}{(\omega+\Omega_m)^2}\sin\Psi(\omega+\Omega_m) \right].\label{dqm}
 \end{equation}
\end{widetext}
 Considering the case of identical RF circuits that are strongly ($\kappa_{1,2} >> \Gamma_{1,2}/\nu_{1,2}$) coupled in such a way so that the difference between the resonant peaks is equal to the resonant frequency of the mechanical oscillator, i.e. $\nu_1=\nu_2=\nu$, $Q_{e1}=Q_{e2}=Q_e$, $\kappa_1=\kappa_2=\kappa=\Omega_m/\nu$, one can show that the maximal additional dissipation factor is inserted into the mechanical oscillator if the RF circuits are pumped at their lower resonant frequency $\omega\simeq \nu/\sqrt{1+\kappa}\simeq\nu-\Omega_m/2$. In this case 
 \begin{equation}
  \Delta Q_m^{-1} \simeq \frac{S U_0^2 \beta^3 \varepsilon_0 Q_e^3}{32 d^3 m \Omega_m^2}.\label{dqm-opt}
 \end{equation}
 If the circuits are pumped at their higher resonant frequency $\omega\simeq\nu/\sqrt{1-\kappa}\simeq\nu+\Omega_m/2$, the additional dissipation factor is negative, with its absolute value still described by Eq. (\ref{dqm-opt}).
 
 In the single-circuit resolved sideband damping regime (when the mechanical mode is coupled with a single-mode electrical resonator with quality factor $Q_e$) the maximal additional dissipation factor is obtained at the pumping frequency $\omega=\nu-\Omega_m$ and can be calculated by setting $\kappa_{1,2}\equiv 0$ in Eqs. (\ref{deltaeff}-\ref{thetaeff}) and (\ref{dqm}). If the condition $Q_e^{-1}\nu<<\Omega_m$ is satisfied, the approximate expression for the maximal additional dissipation factor is  
 \begin{equation}
  \Delta Q_m^{-1} \simeq \frac{S U_0^2 \beta^3 \varepsilon_0 Q_e \nu^2}{16 d^3 m \Omega_m^4}.\label{dqm-resSB}
 \end{equation}
 
 Comparison of (\ref{dqm-opt}) and (\ref{dqm-resSB}) shows that the two-circuit damping scheme allows one to insert dissipation factor that is $(\Omega_m Q_e)^2/(2\nu^2)$ times larger than the maximal dissipation factor that can be obtained in the single-circuit resolved sideband damping regime.
 
 The considered three-mode system can be used for conversion of the vibration amplitude of the mechanical oscillator to the electrical signal. As it follows from equations~(\ref{solform-q1}-\ref{solform-q2}) and~(\ref{sol-q11}-\ref{sol-q11pm}), the amplitudes of the electrical Stokes and anti-Stokes sidebands are proportional to the amplitude of the mechanical vibrations. Enhancement of the transducer response in comparison with the single-circuit resolved sideband regime is obtained due to the simultaneous resonant amplification of the pumping voltage signal at the driving frequency and its Stokes or anti-Stokes sideband.

\section{Experimental setup}
 The interaction between the mechanical oscillator and two inductively coupled electrical RF circuits is studied experimentally. Since the real mechanical oscillator is one of the vibrational modes of the distributed-parameter resonator as well as the capacitive transducer is more complicated than a parallel-plate capacitor, the effective parameters that allow one to transform the real distributed-parameter systems into simplified equivalent lumped-parameter ones are introduced and calculated. This allows us to make a comparison of results of experimental measurements with theoretical predictions obtained in the previous section.  

\begin{figure}
 \includegraphics[width=\columnwidth]{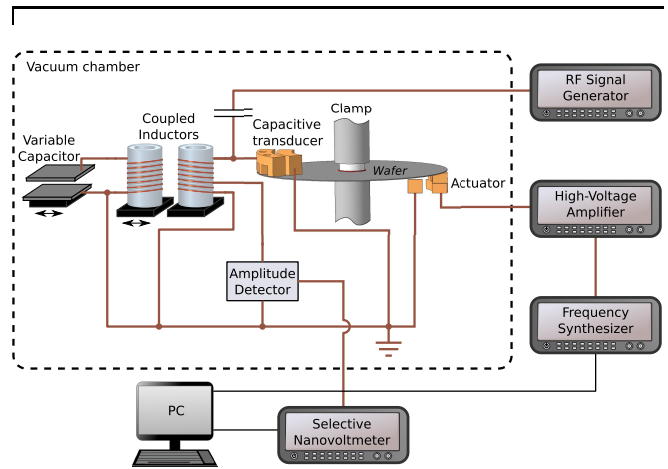}
 \caption{Schematic of the experimental setup.\label{exp-setup}}
\end{figure}

 The mechanical resonator is a double-side polished commercial n-type single-crystal silicon (111) wafer doped with antimony (the electrical resistivity is $0.02$~Ohm$\times$cm). Its diameter is 2a = 76.2 mm, thickness h = 0.34 mm. The wafer is clamped between two 1 cm diameter PTFE stems as it is shown in Fig.~\ref{exp-setup}. The flexural vibration mode (21, 0) with 21 nodal diameters and zero nodal circles is taken because its resonant frequency $\Omega_m=2\pi\times 46478$~Hz and $Q$-factor $Q_m=1.0\times10^{5}$ make it a suitable test mechanical oscillator in the investigated system. The detailed description of such mechanical resonators can be found in Ref.~\onlinecite{Dmitriev2014}. The dynamical behavior of this vibrational mode can be reduced to the lumped mechanical oscillator with the effective mass $m_{eff} = 0.06\times m_{tot} = 2.6\times 10^{-4}$~kg according to the Eq. (4) in Ref.~\onlinecite{Dmitriev2014}.

 The mode vibration is excited by applying an alternating voltage with the frequency equal to a half of the mode resonant frequency to the electrode of the electrostatic actuator that is placed under the wafer as it is shown in Fig.~\ref{exp-setup}.
 
 The comb-shaped electrode of the capacitive transducer which couples the mechanical mode with the RF circuits is placed over the wafer at its opposite to the actuator side. The electrode has 4 lugs with spatial period of 10~mm close to the length of the standing flexural wave excited in the wafer ($\approx 11$~mm). The total area of the surface of the lugs is $S_{tot} = 8.8\times 10^{-5}$~m$^2$. An auxiliary plain electrode is installed near the main electrode closer to the wafer's center so that its surface lies in the same plane with the surface of the lugs. Such a construction allows one to avoid making of a direct electrical contact with the silicon wafer. 
 
 The transducer capacitor in parallel with an inductor $L_1=96$~$\mu$H and a stray capacitor $C_s$ form the first RF circuit that has resonant frequency of about $2.86$~MHz. This circuit is pumped by a RF voltage source which is capacitively coupled to the inductor. The value of the stray capacitance $C_s=13.1$~pF is determined by measuring the resonant frequency of this RF circuit in the absence of the silicon wafer.
 
 The change of the capacitance between the main and auxiliary electrodes caused by the mode vibration can be approximately reduced to the change of the capacitance of the equivalent parallel-plate capacitor with the effective area of the plate $S_{eff}=\int\int_{S_{tot}}w(r,\theta)rdrd\theta\simeq 0.32\times S_{tot}\simeq 2.8\times 10^{-5}$~m$^2$, where $w(r,\theta)$ is the normalized mode shape function. The value of the effective distance between the plates $d_{eff}\simeq 28$~$\mu$m is obtained by finite element modeling of the overall capacitance of this system.
  
 The second RF circuit is formed by an inductor $L_2$ that is identical to the one used in the first circuit and a variable capacitor. The inductor $L_2$ is installed onto a movable platform that allows one to control the mutual inductance of the RF circuits by changing the distance between inductors $L_1$ and $L_2$.
 
 The value of the $Q$-factor of the each of the RF circuits is $Q_{e}=390$.
 
 The amplitude of the wafer vibration is monitored by demodulation of RF signal of the primary circuit using a diode detector which is coupled to the circuit by autotransformer coupling.
 
 Measurements are carried out in vacuum under residual pressure of about $10^{-4}$~Torr.
 
\section{Results and discussion}
 The level of interaction between the mechanical oscillator and the RF circuits can be characterized by the value of the additional dissipation factor $Q_m^{-1}$ inserted into the oscillator due to this interaction. It is calculated as the difference between $Q_m^{-1}$ measured when the primary circuit is pumped by the RF driving voltage with amplitude $U$ and $Q_{m0}^{-1}$ measured when the primary circuit is pumped by RF driving voltage with vanishingly small amplitude. The $Q$-factor of the mechanical oscillator is derived from the measured resonance curve. The curve is fitted to a Lorentzian function which allows calculation of the oscillator resonant frequency and $Q$-factor. In our measurements uncertainties in these values do not exceed 4\%.

 \begin{figure}
 \includegraphics[width=\columnwidth]{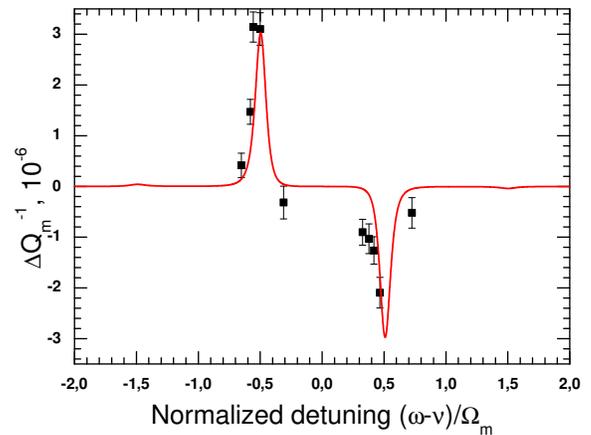}
  \caption{Calculated additional dissipation factor $\Delta Q_m^{-1}$ for the taken vibrational mode of the silicon wafer resonator as a function of normalized detuning $(\omega-\nu)/\Omega_m$ (solid line).  Measured values of the additional dissipation factor (squares).\label{loss-vs-detuning}}
 \end{figure}
 
 According to the theoretical analysis, the maximal additional damping is attained when the difference between the split resonant frequencies of the circuits is equal to the resonant frequency of the mechanical mode. The capacitance of the second RF circuit $C_2$ is tuned so that the amplitudes of voltage across both circuits are equal. In the case of identical inductors this indicates the equality of resonant frequencies of each of the circuits (i.e. $\nu_1=\nu_2=\nu$). The difference between the resonant peaks frequencies is tuned by changing the coupling factor $\kappa$ of the inductors. This difference is maintained equal to the mechanical mode resonant frequency. The calculated additional mechanical dissipation factor $\Delta Q_m^{-1}$ as a function of dimensionless detuning $(\omega-\nu)/\Omega_m$ is shown by solid line in Fig.~\ref{loss-vs-detuning}. The negative value of this factor corresponds to the regenerative effect. Measured values of the additional mechanical dissipation factor for the taken 
vibrational 
mode of the silicon wafer resonator are marked by the squares.

 \begin{figure}
  \includegraphics[width=\columnwidth]{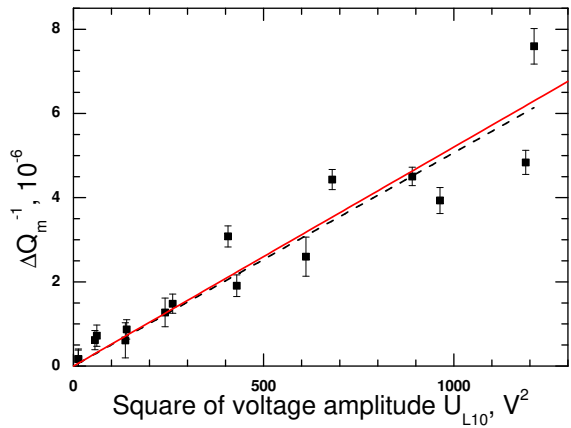}
  \caption{The additional dissipation factor $\Delta Q_m^{-1}$ as a function of square of the voltage amplitude across the inductor $L_1$ calculated (solid line) and measured (squares) for the optimal tuning of the three–mode system. The dashed line represents the linear fit of the experimental data.\label{loss-vs-sqvoltage}}
 \end{figure}
 
 When the frequency of the RF signal generator pumping the circuits is set equal to the lower one of the two split resonant frequencies of the system, the additional dissipation factor becomes positive. This corresponds to an increase of the full damping of the mechanical mode. When the frequency of the RF signal generator is set equal to the higher frequency the additional dissipation factor turns negative. This corresponds to a decrease of the full damping of the mechanical mode.  Fig.~\ref{loss-vs-sqvoltage} shows the dependence of the modulus of the additional mechanical dissipation factor on the square of the RF voltage amplitude across the inductors, $U_{L_{10}}=U_{L_{20}}\simeq U_0 Q_e/2$. The same dependence calculated in accordance with Eq.~(\ref{dqm-opt}) is also drawn in Fig.~\ref{loss-vs-sqvoltage} as a solid line. One can find a good agreement between the calculated and measured values.

\section{Conclusion}
 An electromechanical system that consisted of a high-$Q$ mechanical oscillator parametrically coupled in the manner of a capacitive transducer with a RF circuit, which is in turn inductively coupled with another RF circuit, has been studied both theoretically and experimentally. The primary and secondary RF circuits are nearly identical. The coupling between circuits is strong, i.e. the two split resonant peaks are resolved due to the fulfillment of the condition $\kappa >> Q_{e1,2}^{-1}$, where $Q_{e1,2}$ are the quality factors of the two RF circuits. One of the flexural vibration modes of a circular silicon wafer has been used as a mechanical oscillator. The vibrations of the wafer have modulated the RF electric current in the primary circuit, producing two frequency sidebands known as Stokes and anti-Stokes ones. The coupling factor between the two circuits has been set equal to the 
ratio of the mechanical mode frequency to the resonant frequency of each of the circuits ($\kappa=\Omega_m/\nu$), so that the difference between the split resonant frequencies of the coupled RF circuits has been equal to the mechanical mode frequency. This allowed the carrier and one of the sideband frequencies to be amplified simultaneously, significantly enhancing the effect of the interaction between the mechanical mode and the electrical circuits in the so-called resolved sideband regime. Notice that this behavior is the result of parametric coupling between the linear mechanical and electrical systems. The electrostatic force in the capacitor produces back action on the wafer and inserts additional stiffness and dissipation factor that can be positive or negative. So the nonlinearity of the dependence of the electrostatic force on the charge of the capacitor plates is essential for this effect. If the driving frequency is equal to the lower of the electrical circuit split frequencies, the additional 
damping is positive. If the driving frequency 
is equal to the upper one, the additional damping is negative. Measurements of the additional mechanical dissipation factors which have been carried out using the experimental setup have shown good agreements with the calculated ones. Such a coupled electromechanical system can be used as a high-sensitive capacitive vibration sensor.
\begin{acknowledgments}
The authors gratefully acknowledge the support of the Russian Foundation For Basic Research under grant 14-02-00399 and the United States National Science Foundation and Caltech under grant PHY-1305863 as well as the invaluable help of Stan Whitcomb.
\end{acknowledgments}

\end{document}